\documentclass[a4paper,11pt]{article}

\usepackage{amsmath}
\usepackage{graphicx}
\usepackage[hyphens]{url}
\usepackage[hypertexnames=false,breaklinks=true]{hyperref}
\usepackage{booktabs}
\usepackage{xcolor}
\usepackage[margin=1in]{geometry}
\usepackage{authblk}
\usepackage{float}
\usepackage[comma,square,numbers,sort&compress]{natbib}
\usepackage{adjustbox}

\title{The Evolution of Heavy-Ion Physics: A Data-Driven Analysis of Quark Matter Conferences}

\author[1]{D.J.~Kim}
\author[2]{T. Lappeteläinen}

\affil[1]{Department of Physics, University of jyväskylä, Finland}

\begin{document}

\maketitle

\begin{abstract}
This paper presents a data-driven analysis of Quark Matter conferences from 2011 to 2025, investigating trends in geographical representation, research emphasis, and methodological strategies. Using a dataset of over 10,000 presentations, the evolution of heavy-ion physics through its premier conference series is examined. The distribution of presentations across countries and institutions shows the global reach of the field while highlighting opportunities for broader international engagement. Research topics show clear temporal trends, with certain physics concepts rising and falling in prominence, and a gradual shift from facility-focused to phenomenon-focused research. The analysis reveals increasing integration between theoretical and experimental approaches over time, reflecting a maturing field where these complementary domains strengthen and enhance each other. These findings provide quantitative insights that the evolution of heavy-ion physics is strengthened by international collaboration at conferences. The analytical methods developed here could help other scientific communities understand their own patterns of knowledge sharing and development.
\end{abstract}

\section{Introduction}
The Quark Matter conference series represents the premier international meeting dedicated to ultra-relativistic heavy-ion collisions and studying nuclear matter under extreme conditions. Since its inception in 1980 at Berkeley, USA, as a workshop on ultra-relativistic heavy-ion collisions, these conferences have served as crucial platforms for presenting breakthrough research, fostering collaborations, and shaping the direction of the field. The conference has grown substantially over the decades, rotating between North America, Europe, and Asia approximately every one or two years, reflecting the truly international character of heavy-ion physics research. A complete list of all conference proceedings from 1980 to 2025 is provided in Appendix~\ref{app:proceedings}, providing a historical record of the field's evolution.

\begin{figure}[H]
\centering
\includegraphics[width=\textwidth]{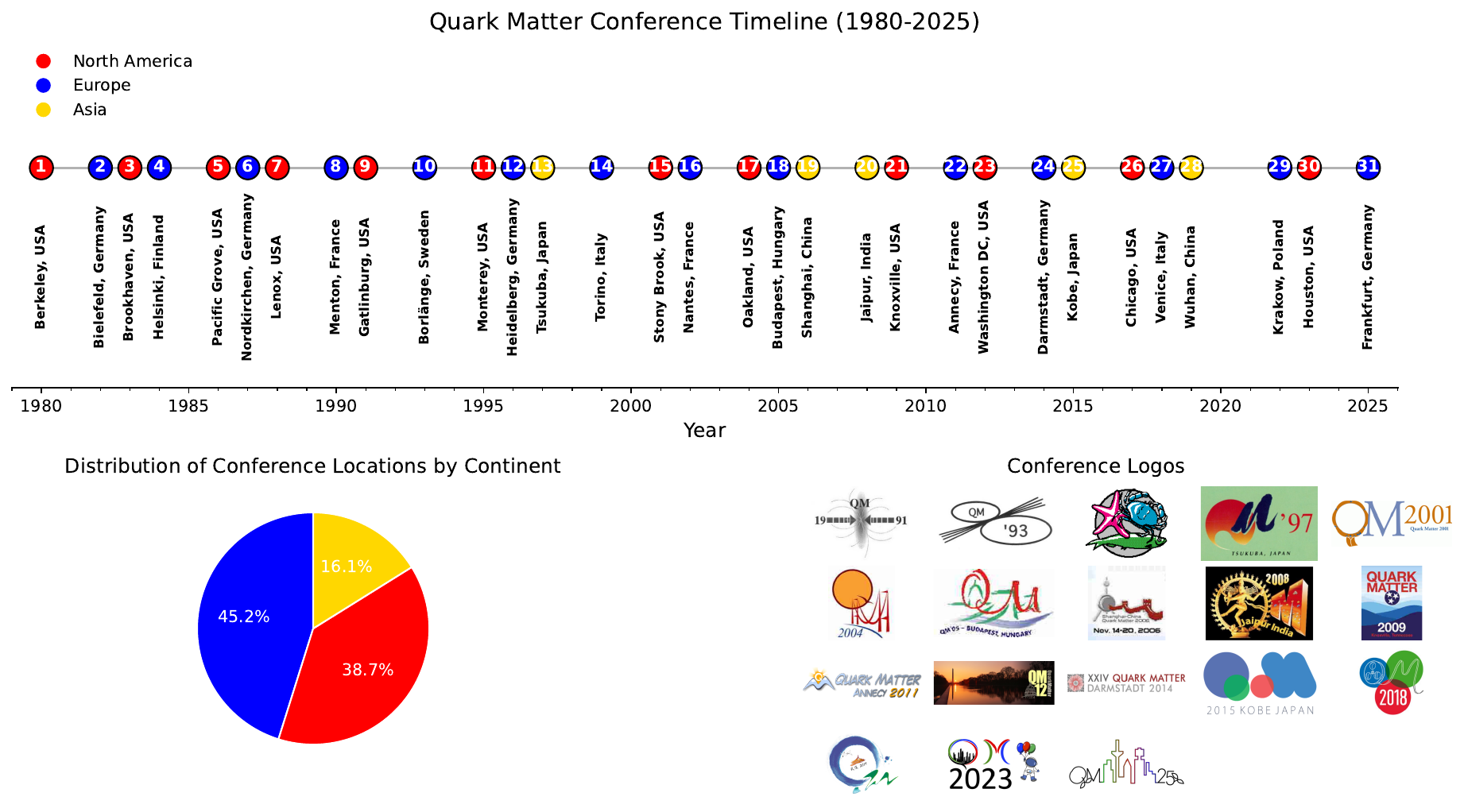}
\caption{Historical timeline of Quark Matter conferences from 1980 to 2025. The top panel shows the chronological progression of all 31 conferences with their locations, color-coded by continent (blue for Europe, red for North America, and gold for Asia). The bottom panel displays the continental distribution of conference venues.}
\label{fig:venues_all}
\end{figure}

Figure~\ref{fig:venues_all} shows the whole historical record of Quark Matter conferences from its founding in 1980 through the 2025 session. The conference began as a relatively small workshop at Berkeley and has since expanded into a major international event with rotating venues across three continents. Although Asian locations have been introduced at less regular intervals, accounting for 16.1\% of all conferences, the geographical distribution demonstrates an intentional balance between North America and Europe, which together have held 83.9\% of all conferences. This distribution shows the worldwide footprint of the field gradually expanding, as well as the historical centers of heavy-ion physics study. This distribution reflects the historical centers of heavy-ion physics research and the gradual expansion of the field's global footprint.
The timeline also shows interesting temporal patterns in venue selection. In the early years (1980-1990), conferences were held more frequently, sometimes annually, as the field was rapidly developing. The pattern later stabilized to a roughly biennial schedule, with occasional adjustments. Notable disruptions include the COVID-19 pandemic period, which affected the scheduling between the 2019 Wuhan conference and the 2022 Krakow event. The consistent rotation between continents demonstrates the community's commitment to international accessibility, ensuring that researchers from different regions have opportunities to participate without persistent geographical barriers.

Heavy-ion physics investigates the behavior of nuclear matter at extreme temperatures and densities where protons and neutrons dissolve into their constituent quarks and gluons, forming a state of matter known as the Quark-Gluon Plasma (QGP)~\cite{Shuryak:2008eq, Kovtun:2004de}. This deconfined state of matter is believed to have existed microseconds after the Big Bang and may currently exist in the cores of neutron stars. Particle accelerators such as RHIC at Brookhaven in US and the LHC at CERN in Europe enable physicists to replicate extreme conditions via high-energy collisions of heavy nuclei, facilitating laboratory investigations of QGP properties~\cite{Adams:2005dq, Adcox:2004mh, Arsene:2004fa, Back:2004je, ALICE:2022wpn}. The field bridges particle physics, nuclear physics, and astrophysics, addressing fundamental questions about quantum chromodynamics (QCD), phase transitions in nuclear matter, and collective phenomena in strongly interacting systems~\cite{Busza:2018rrf,Brambilla:2014jmp,Fukushima:2010bq}. These investigations enhance our understanding of the early universe, the strong nuclear force, and the fundamental structure of matter under extreme conditions~\cite{Akiba:2015jwa}.

In recent years, the field has achieved remarkable progress in quantitatively characterizing QGP properties. The transport coefficients of QGP, including shear viscosity, bulk viscosity, and specific heat, have been determined with unprecedented precision through global Bayesian analysis of experimental data from RHIC and the LHC integrated with state-of-the-art theoretical models~\cite{Bernhard:2016tnd, JETSCAPE:2020mzn, Nijs:2020roc}. Current research frontiers include understanding how energetic probes like jets interact with the medium (e.g. \cite{JETSCAPE:2020mzn}), elucidating the behavior of heavy flavor quarks (charm and beauty) as they propagate through QGP (e.g.~\cite{Rapp:2018qla}), and exploring the transition between normal nuclear matter and QGP through beam energy scan programs (e.g.~\cite{Busza:2018rrf,Brambilla:2014jmp,Fukushima:2010bq,Giacalone:2023cet}). These investigations aim to establish a framework describing the time evolution, thermodynamic properties, and microscopic dynamics of strongly interacting QCD matter~\cite{Bernhard:2015hxa,Bernhard:2016bar,Bernhard:2016tnd,Bernhard:2018hnz,Bernhard2019,Auvinen:2020mpc,Nijs:2020ors,Nijs:2020roc,JETSCAPE:2020mzn,Parkkila:2021tqq, Parkkila:2021yha, Virta:2024avu}.

The future of heavy-ion physics at the LHC holds significant promise with planned upgrades and extended run periods. The upcoming Run 3 and Run 4 at the LHC will provide substantially larger data samples, enabling more precise measurements of rare probes and allowing for multi-differential analyses that were previously statistically limited~\cite{ALICE:2022wpn,Citron:2018lsq,Apollinari:2017lan,Bruce:2022qrm,ALICE:2022wwr,CMS:2022ixl,ATLAS:2022hro}. The High-Luminosity LHC (HL-LHC) upgrade, scheduled to begin operations in the 2030s, will further increase collision rates by a factor of 5-7 compared to the LHC's design values, opening new possibilities for ultra-rare measurements~\cite{Brewer:2022vkq}. Major detector upgrades for ALICE, ATLAS, and CMS will enhance capabilities for heavy-ion measurements, with improvements in tracking resolution, particle identification, and data acquisition rates~\cite{CMS:2022ixl,ATLAS:2022hro,ALICE:2022wwr}. These technological advances will enable unprecedented studies of jet quenching, heavy flavor production, and electromagnetic probes of the QGP. Additionally, proposals for future facilities like the electron-ion collider (EIC) in the United States and potential heavy-ion programs at FCC will extend the field's reach in coming decades, promising new insights into QCD matter under extreme conditions~\cite{Akiba:2015jwa,Citron:2018lsq}.

Understanding these conferences' historical patterns and evolution provides valuable insights into the landscape of research topics and methodologies over time. It illuminates geographical and institutional representation patterns in the field, showing both centers of research activity and potential gaps in participation. Such analysis highlights opportunities to show diversity and inclusivity in scientific participation, which is essential for the field's continued success. Additionally, it helps track the impact of major experimental programs and theoretical advances on research directions and community focus.

In this paper, we analyze data from Quark Matter conferences spanning from 2011 to 2025, extracting patterns from presentation titles, speaker affiliations, and presentation categories to provide a quantitative assessment of how the field has evolved. Our analysis encompasses over 10,000 presentations across multiple conference cycles.
The paper is organized as follows: Section~\ref{sec:data_and_methods} describes the data collection and analytical approaches. Section~\ref{sec:results} presents the results, including the geographical distribution of contributions, research topic trends, and representation patterns. Section~\ref{sec:discussion} discusses these findings.

\section{Data and Methods}
\label{sec:data_and_methods}
Data are collected from the official Indico pages of Quark Matter conferences from 2011 to 2025 (see the complete list in Appendix~\ref{app:indico}). For each conference, we extracted presentation titles, authors, and affiliations, along with presentation categories (plenary, parallel, poster) and information about the location and year of each event. This dataset provided the foundation for our subsequent analyses of participation patterns and research trends. Our study focuses on this recent period (2011-2025) because it represents the era when conference data became available through the Indico platform, allowing for data collection and analysis. The earlier conferences used different documentation systems and inconsistent data formats, which will require substantial work, and we will leave it for future work. Additionally, this recent period coincides with significant developments in the field, including the operation of the LHC. Major upgrades at RHIC provide a coherent timeframe for examining how these experimental advances have influenced research directions.

Two complementary Python scripts were developed for data collection and analysis. The first step fetches raw data from Indico pages using conference IDs specified in a reference file. The second step is to process the collected data to generate statistics and visualizations. This separation allows for more efficient data processing, with the first step handling the time-consuming task of fetching data from web sources. The latter analysis script focuses on analysis and visualization.
Our analysis pipeline included classifying presentations into plenary, parallel, and poster categories, extraction of speaker information including names, institutes, and countries, text processing of presentation titles to identify key research topics, and aggregation of statistics by conference, institute, and country.

An enhancement to our study was developing an affiliation resolution system. This system utilized a multi-stage approach to country detection from affiliation strings, applied pattern matching using a database of institution keywords, and cross-referenced against an extensive mapping of institutions to countries. It also handled name variations and formatting inconsistencies to maximize matching, including country code detection (e.g., "(US)," "(DE)") and explicit country name matching. Because of the different handling of the affiliation strings, the country detection system is not fully automated and requires manual review and correction.
This methodology substantially reduced the number of speakers with unknown affiliations from over 15\% in the initial dataset to less than 3\% in the final analysis. The remaining cases were manually reviewed and, where possible, corrected by referring to publication records and institutional websites.

As for viewing the research highlight of the conference series, we extracted keywords from presentation titles after removing common stop words and non-meaningful terms. These extracted keywords in each conference event allowed us to track the evolution of research focus over time.

\section{Results}
\label{sec:results}
\subsection{Conference Venues and Geographical Scope}

\begin{figure}[H]
\centering
\includegraphics[width=\textwidth]{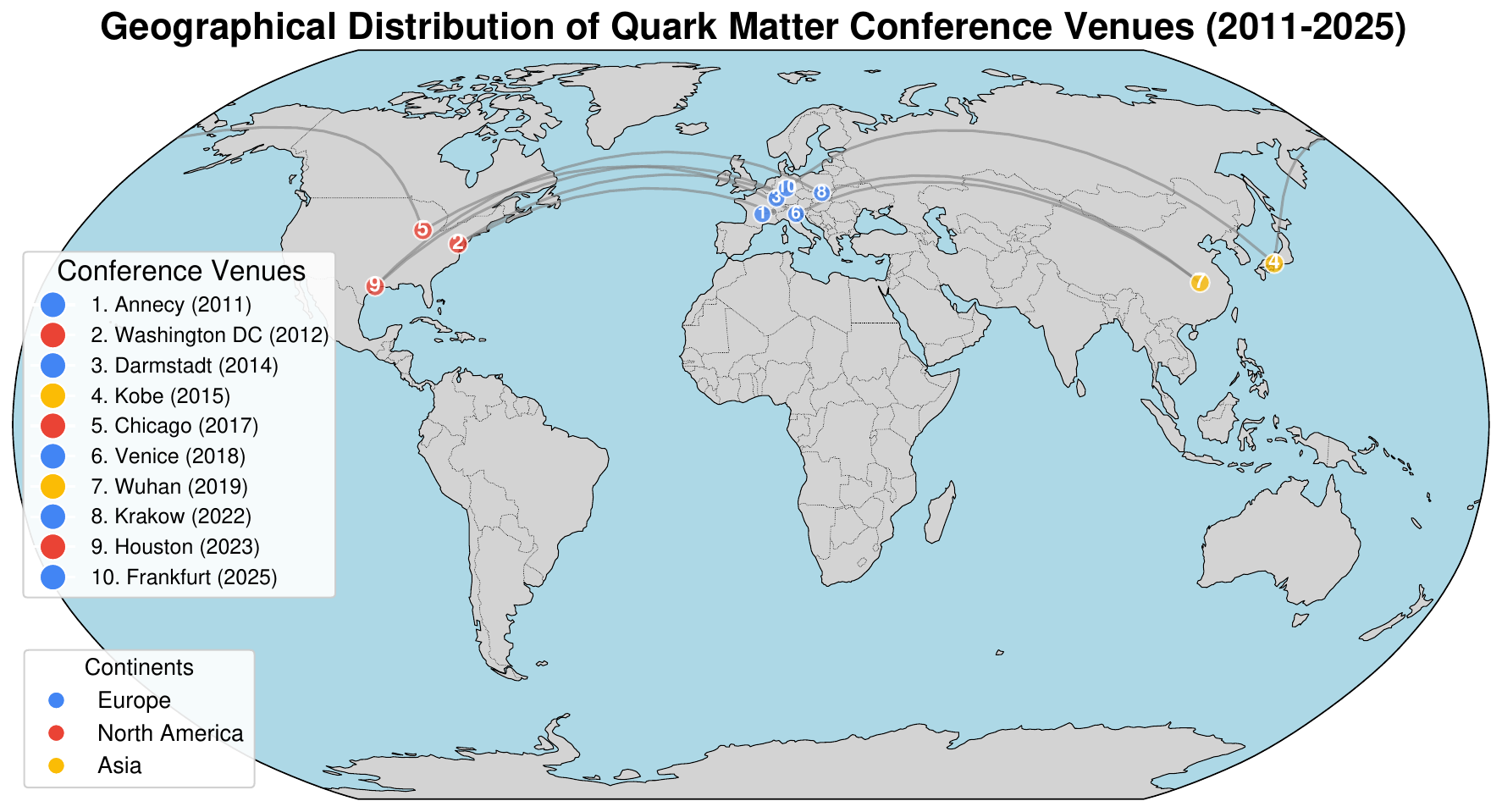}
\caption{Geographical distribution of Quark Matter conference venues from 2011 to 2025. The map shows the global spread of conference locations across three continents, with numbered markers indicating the chronological sequence. Venues are color-coded by continent (see the legend at the top left).}
\label{fig:venues}
\end{figure}

Figure~\ref{fig:venues} presents the geographical distribution of Quark Matter conference venues over time. The venue selection shows an organized effort to alternate the conference across different global regions, with representation from North America, Europe, and Asia throughout the analyzed period. This rotation approach helps ensure that the conference remains accessible to researchers worldwide over time, even if individual conferences may have geographical attendance biases.
There appears to be a pattern of alternating continents, particularly between Europe and North America, with Asian venues interspersed at less regular intervals. This pattern reflects the historical centers of heavy-ion physics research and efforts to expand the conference's international footprint.
The frequency of venues in different regions roughly corresponds to the size of the heavy-ion physics community in those areas, with Europe and North America hosting most frequently, followed by Asia.

The choice of venue correlates with participation patterns, as shown in our analyses of speaker institutes. Conference attendance is typically higher in the host country and region, affecting both the volume and diversity of submissions from different geographical areas.
As shown in the historical timeline (Figure~\ref{fig:venues_all}), this pattern of continental rotation has been a consistent feature of the Quark Matter conference series since its beginning, reflecting the field's commitment to international accessibility and collaboration. The recent period analyzed in this study (2011-2025) continues this tradition while showing a gradual expansion in the geographical diversity of participants, even as venue selection remains concentrated in traditional research centers.

\subsection{Conference Statistics and Participation Trends}

\begin{figure}[H]
\centering
\includegraphics[width=\textwidth]{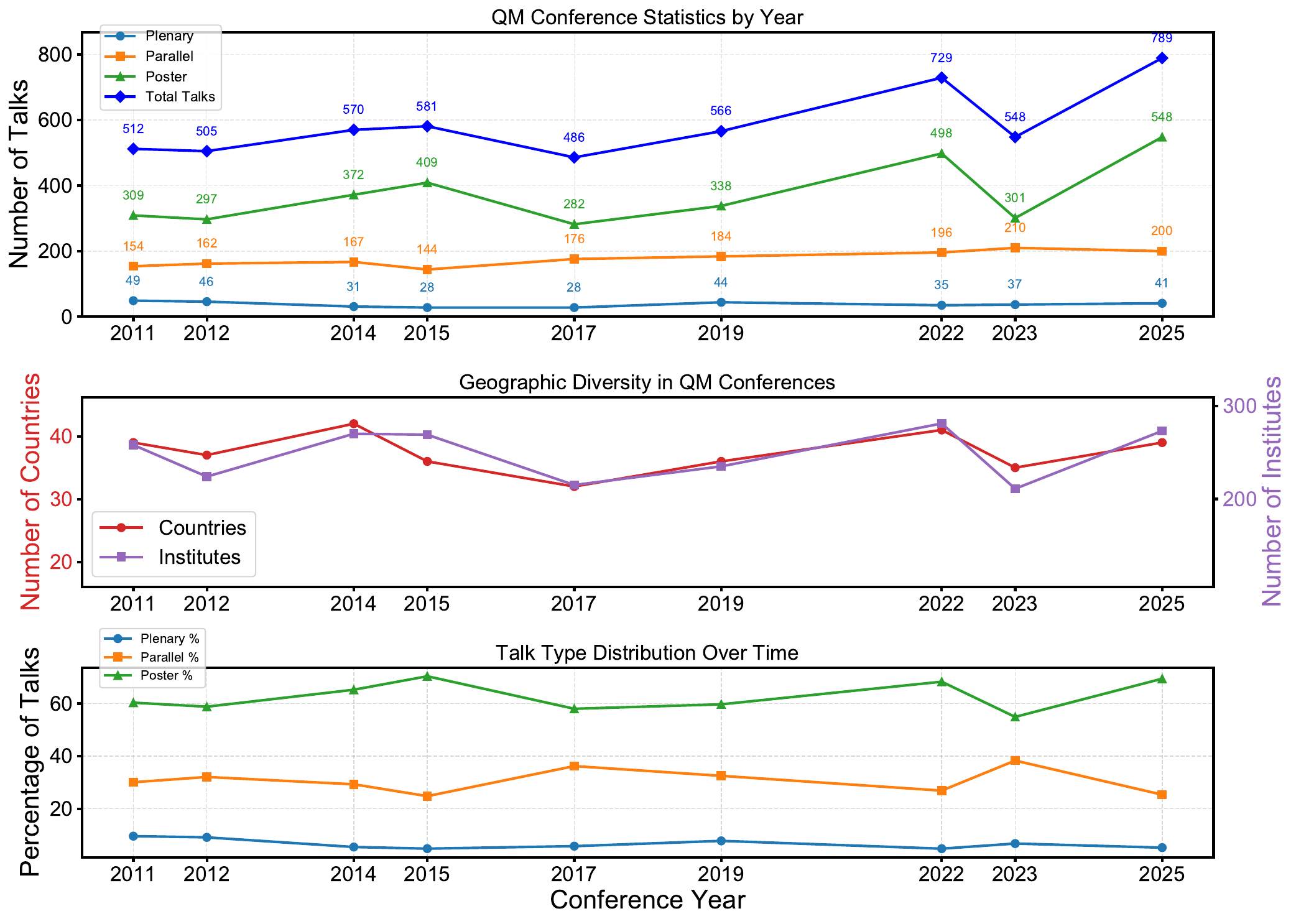}
\caption{Statistical overview of Quark Matter conferences from 2011 to 2025. The top panel shows the number of participants and presentations of various types (plenary, parallel, poster) across conference years. The middle panel tracks the number of countries and institutes represented at each conference. The bottom panel displays the percentage distribution of talk types over time.}
\label{fig:talk_statistics}
\end{figure}

Figure~\ref{fig:talk_statistics} provides a statistical overview of Quark Matter conferences over the analyzed period.
The total number of presentations varies between conferences, reflecting both changes in the size of the heavy-ion physics community and the practical constraints of different venues. This variation affects the competitiveness of selection processes, particularly for high-visibility plenary and parallel talks.
The overall participant attendance fluctuates over the years, with specific conferences attracting significantly larger audiences than others. Notably, the significantly lower attendance in 2022 can be attributed to the COVID-19 pandemic, which continued to impact international travel and large gatherings. This conference was held during a transition period from virtual to in-person events, with many institutions still maintaining travel restrictions. In some cases, the number of participants is lower than the total number of contributions, which is made possible through remote presentations via Zoom and similar platforms, allowing researchers to present their work without physically attending the conference. 

The selection process for contributions also influences participation patterns. Quark Matter conferences typically employ a specific abstract review process, where submissions are evaluated by the program committee based on scientific merit and relevance. While plenary and parallel talks undergo rigorous selection, poster acceptance rates are generally higher, creating different participation opportunities. Additionally, institutional funding policies often play a role in attendance decisions, as many research institutions only provide travel funding for researchers who have an accepted contribution. This funding constraint can create a feedback loop where researchers from less-represented institutions find participating difficult.
Excluding this pandemic-affected outlier, participant numbers do not always directly correlate with the total number of presentations, suggesting that factors such as conference location, accessibility, concurrent events, and broader community interest play important roles in determining attendance. The highest attendance periods often coincide with conferences held in major research hubs or following significant experimental milestones.

The ratio between different presentation types (plenary, parallel, poster) has evolved, as shown in the bottom panel. This distribution reveals shifts in conference organization strategies and the relative emphasis on different presentation formats. The percentage analysis shows that while plenary talks typically constitute a small fraction of total presentations (around 5-10\%), their proportion has remained stable. The balance between parallel and poster presentations shows more variation, likely reflecting both venue constraints and deliberate choices by conference organizers.

The middle panel tracks geographical diversity (number of countries represented) and institutional diversity (number of institutes). The number of participating countries has ranged from 22 to 30 across conferences, while the number of represented institutes has varied between 230 and 310 for most years. The overall trend shows an increase in both metrics, suggesting the conference has become more internationally diverse over time, with some fluctuations correlating with conference locations.
As explored in our subsequent analyses, these structural patterns in conference organization and participation have implications for visibility distribution across different research groups and geographical regions.

Even though it is not shown in the figures, flash talks have been a consistent feature of Quark Matter conferences since 2011, providing selected poster presenters an opportunity to share their work with a broader audience. Table~\ref{tab:flash_talks} shows the number of flash talks at each conference over the analyzed period. This format offers multiple benefits to the community: it creates a stepping stone for early-career researchers to gain speaking experience; it increases visibility for work that might otherwise receive limited attention in poster sessions; it provides a mechanism for highlighting clever approaches from smaller research groups; and it adds programming flexibility for organizers to feature emerging topics. Additionally, from participation in these conferences, it has been observed that flash talks serve as an effective talent-spotting mechanism, with several past flash talk presenters later becoming invited speakers at subsequent conferences. The format has proven valuable for increasing the visibility of interdisciplinary work that might not fit into traditional parallel session categories. This implementation is a good example of how thoughtful conference organizations can enhance scientific exchange and community development.

\begin{table}[H]
\centering
\caption{Number of flash talks at Quark Matter conferences (2011-2025)}
\label{tab:flash_talks}
\begin{tabular}{c|cccccccccc}
\toprule
\textbf{Year} & 2011 & 2012 & 2014 & 2015 & 2017 & 2018 & 2019 & 2022 & 2023 & 2025\\
\midrule
\textbf{Talks} & 8 & 8 & 8 & 8 & 8 & 10 & 6 & 10 & 10 & 11\\
\bottomrule
\end{tabular}
\end{table}

This format has maintained a consistent presence, with 8 flash talks in each conference from 2011 to 2017, a slight increase to 10 in 2018, a temporary decrease to 6 in 2019, and a return to 10 flash talks in both 2022 and 2023. Flash talks represent a very small fraction of poster presentations, creating an intermediate visibility tier between standard posters and parallel sessions. These presentations have been particularly beneficial for researchers from institutions with less established presence at Quark Matter, where the opportunity enhanced their conference experience and professional networking.

\subsection{Research Topic Evolution}

\begin{figure}[H]
\centering
\includegraphics[width=\textwidth]{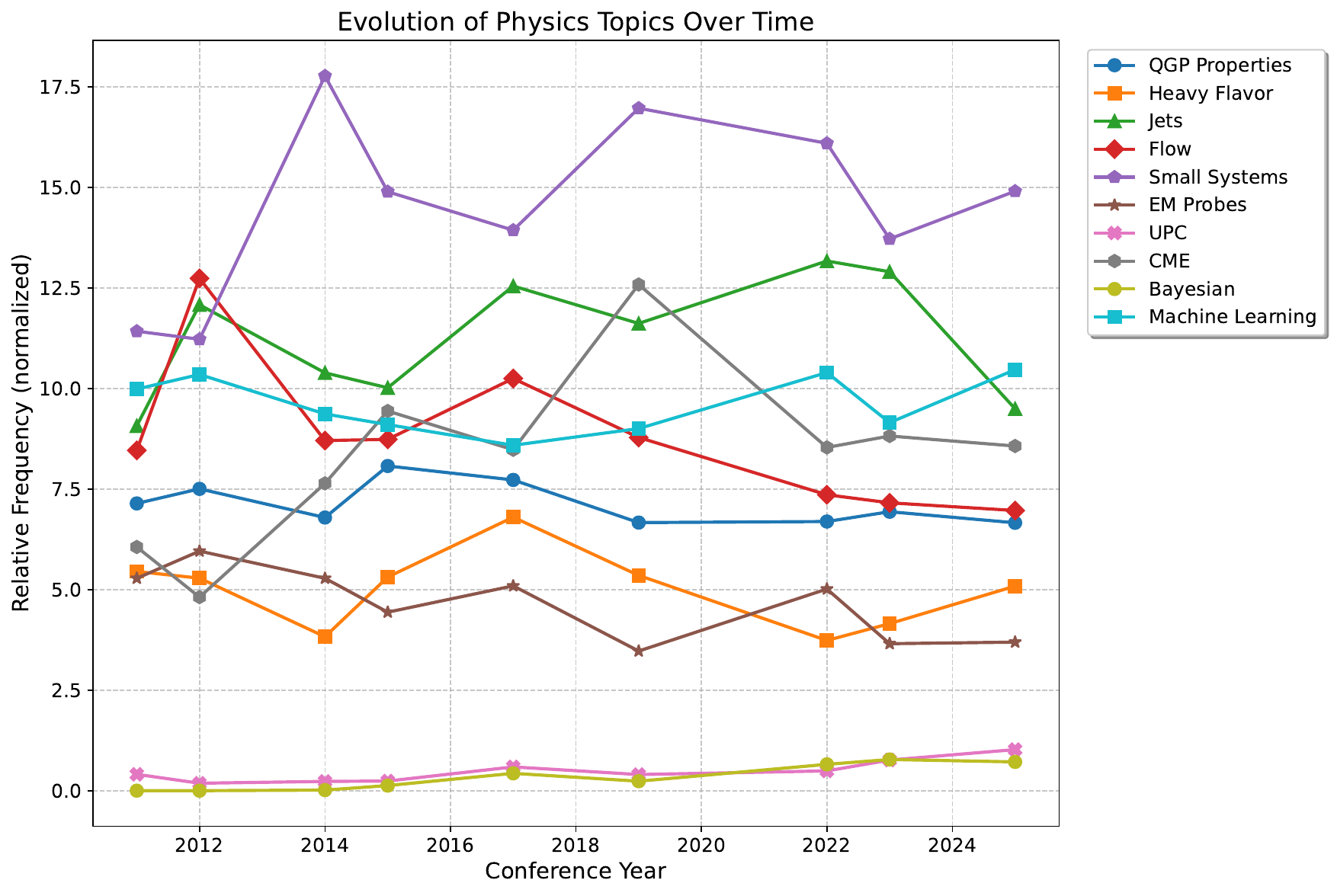}
\caption{Analysis of research topics in Quark Matter conferences from 2011 to 2025. The figure shows the evolution of key physics concepts as relative frequencies over time. This shows quantitative trends in research focus over time, tracking the rise and fall of specific physics topics such as QGP properties, flow phenomena, small systems, and emerging areas like machine learning applications.}
\label{fig:keywords}
\end{figure}

Figure~\ref{fig:keywords} presents an analysis of research topics across Quark Matter conferences from 2011 to 2025. This tracks specific physics concepts over time, revealing clear temporal patterns in research focus. "QGP" terms maintain a consistent presence throughout the period, confirming the central role of quark-gluon plasma studies in the field. "Flow" (e.g.~\cite{Heinz:2013th}) shows peak interest in the middle years (2014-2018), while "Small Systems" (e.g.~\cite{Loizides:2016tew}) and "UPC" (ultra-peripheral collisions, e.g.~\cite{Bertulani:2005ru,Bertulani:2024hmb}) gain prominence in later conferences, reflecting the community's expanding research frontiers. "Jets" show a relatively stable presence with some fluctuations, highlighting their enduring importance as probes of the QGP medium~\cite{Cao:2022odi}. The tracking of "HF" (Heavy Flavor) demonstrates the evolution of charm and beauty quark physics throughout the period~\cite{Zhang:2023oid,Xing:2021bsc,Xing:2019xae}. The emergence of "Machine Learning" applications (see e.g.~\cite{Pang:2020ipp}) and "Bayesian" analysis techniques~\cite{Bernhard:2015hxa,Bernhard:2016bar,Bernhard:2016tnd,Bernhard:2018hnz,Bernhard2019,Auvinen:2020mpc,Nijs:2020ors,Nijs:2020roc,JETSCAPE:2020mzn,Parkkila:2021tqq, Parkkila:2021yha, Virta:2024avu} in recent years highlights the field's adoption of advanced computational and statistical methods for data interpretation. It should be noted that detecting correlations between different physics domains (such as jets and flow) presents analytical challenges due to the complex terminology used across these interconnected research areas.

\begin{figure}[H]
    \centering
    \includegraphics[width=\textwidth]{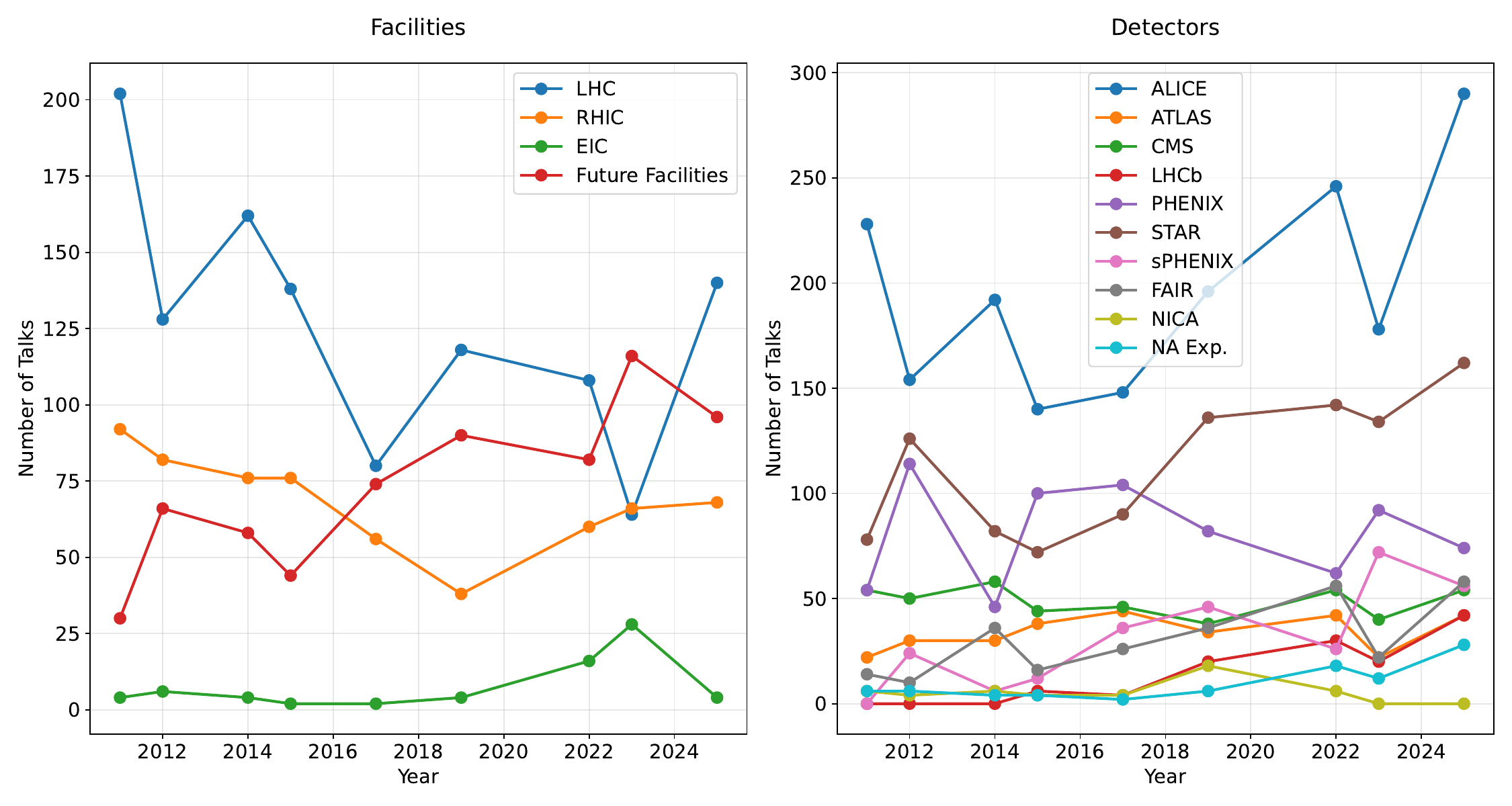}
    \caption{Trends in experimental facilities mentions at Quark Matter conferences from 2011 to 2025. The figure displays the number of mentions for different detectors and experimental facilities over time.}
    \label{fig:detector_focus}
\end{figure}

Figure~\ref{fig:detector_focus} shows the experimental facilities across the same period, illustrating the field's transition from facility-focused to phenomenon-focused research. Early conferences show strong emphasis on LHC experiments (ALICE, ATLAS, CMS) and RHIC facilities (STAR, PHENIX), with some fluctuation in detector mentions over time. The increasing mentions of future facilities in recent years (sPHENIX, EIC, NICA) signals the field's forward-looking perspective and preparation for next-generation experiments. This reveals how the community's attention shifts between different experimental programs as new capabilities come online and research priorities evolve.

These figures together reveal shifting priorities in both physics topics and experimental approaches. For instance, we can observe the emergence of topics like Chiral Magnetic Effect (CME)~\cite{Kharzeev:2015znc} in Figure~\ref{fig:keywords} alongside the changing emphasis on different detector collaborations in Figure~\ref{fig:detector_focus}, providing complementary views of how the field has evolved over the past decade.
The temporal trends also highlight the impact of major experimental milestones on research focus. Peaks in certain topics often follow significant data-taking periods at RHIC or the LHC, demonstrating how the field responds to new experimental capabilities and results.
This also captures emerging research frontiers that have gained momentum in recent years. Topics such as small system collective effects, which investigate whether QGP-like behavior occurs in proton-nucleus collisions, show notable growth trends that suggest developing research directions.
By tracking the relative prominence of different topics over time, this analysis provides a quantitative basis for understanding how the landscape of heavy-ion physics has evolved. This information can help researchers identify both established areas with sustained interest and opportunities for novel contributions.

\subsection{Theory vs. Experimental Balance}

\begin{figure}[H]
\centering
\includegraphics[width=\textwidth]{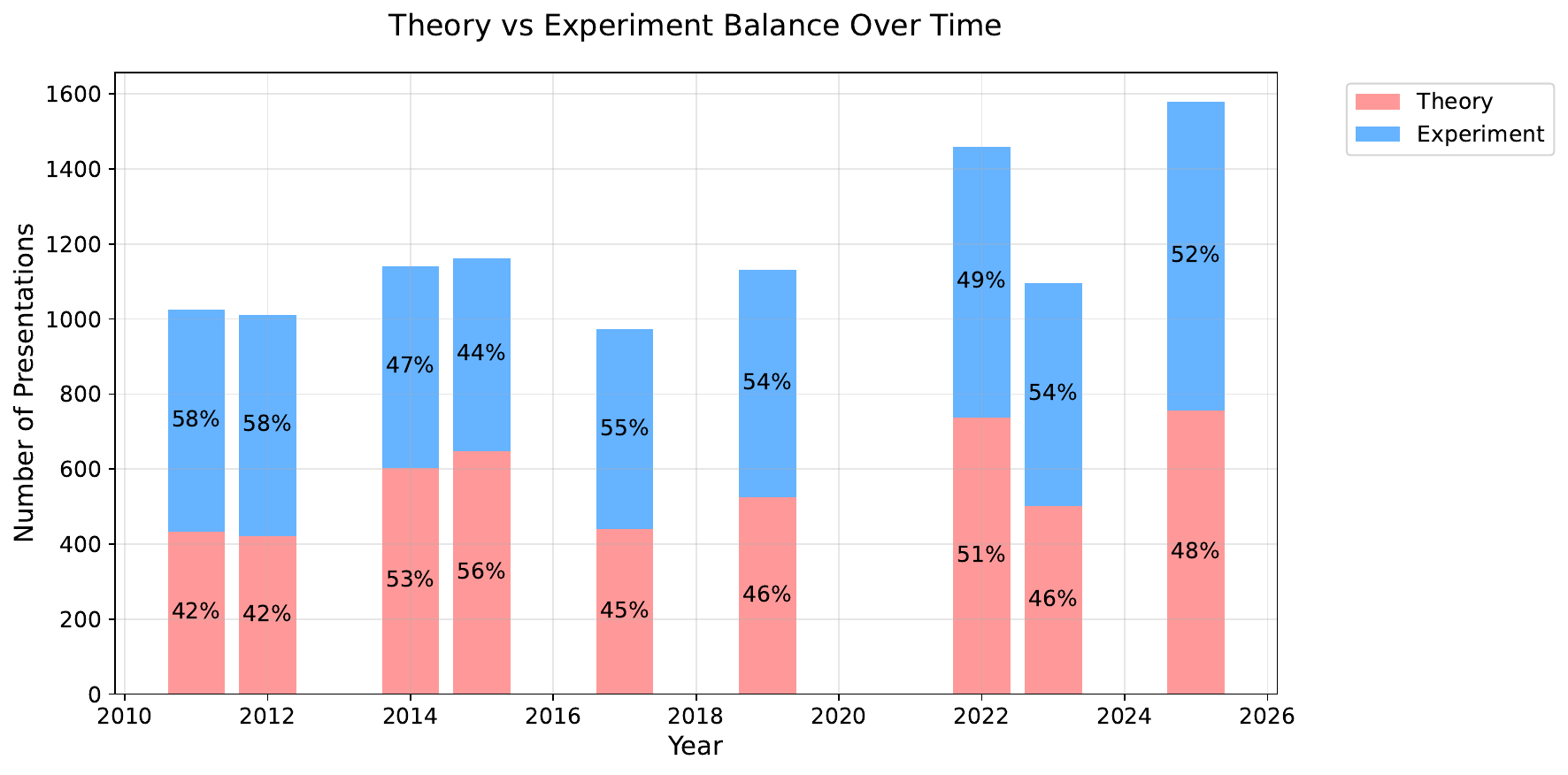}
\caption{Evolution of theory versus experimental contributions in Quark Matter conferences from 2011 to 2025.}
\label{fig:theory_experiment}
\end{figure}

Figure~\ref{fig:theory_experiment} shows the balance between theoretical and experimental contributions across Quark Matter conferences. The classification is done as follows: if a presentation's title contains specific experiment or detector names (e.g., ALICE, ATLAS, CMS, STAR, PHENIX), it is categorized as experimental; otherwise, it is classified as theoretical. This classification approach reflects the conference's standard practice, where experimental talks typically reference their specific detector or collaboration in the title.

The data reveal that experimental contributions typically comprise between 44\% and 58\% of all presentations, with an average of around 52\%. Theoretical presentations maintain a substantial presence, ranging from 42\% to 56\% over the years. Notably, during 2014-2015, theoretical contributions briefly became the majority (53-56\%). The most recent conferences (2022-2025) show a balanced distribution, with experimental presentations only slightly predominant at 49-54\%. These fluctuations in the theory-experiment ratio provide valuable insight into how the field responds to and integrates new experimental capabilities with theoretical interpretations and predictions, particularly following major experimental milestones such as new LHC runs or detector upgrades.

This suggests a well-established field where both approaches maintain their important roles. While the ratio shows some variation in response to experimental milestones, the consistent presence of theoretical and experimental work, typically in a roughly 45-55 split, reflects a field where theory and experiment are integrated. This integration is particularly evident in how theoretical frameworks are regularly tested against new experimental data and how experimental observations drive theoretical developments, creating a productive dialogue that advances our understanding of heavy-ion physics.

These integrated works suggest a maturing field in which theoretical and experimental approaches increasingly inform each other rather than developing in parallel. This trend is further reflected in recent funding patterns, where collaborative projects that bridge theory and experiment have seen increased success rates. 
Notable examples include the JETSCAPE collaboration, which has secured significant multi-year funding to develop a framework integrating theoretical models with experimental constraints~\cite{JETSCAPE:2020mzn}, and the Center of Excellence in Quark Matter, which has successfully obtained substantial funding by emphasizing the synergy between theoretical development and experimental validation~\cite{CoEQM2022}. The Center of Excellence in Quark Matter, funded by the Research Council of Finland, explicitly combines "three theoretical and two experimental research teams" to address questions about quark-gluon plasma formation, demonstrating how funding agencies prioritize integrated approaches~\cite{CoEQM2022}.

Other successful examples of theory-experiment integration include the U.S. Department of Energy's Topical Collaboration on "Jet and Electromagnetic Tomography of Extreme Phases of Matter in Heavy-ion Collisions" (JET Collaboration), which received renewed funding based on its success in bringing together theorists and experimentalists to develop a quantitative understanding of jet quenching~\cite{JET:2013cls}. Similarly, the European Research Council has funded several ERC grants that explicitly bridge theoretical and experimental approaches, such as the "STRONGINT" project, which combines theoretical nuclear physics with experimental constraints to understand the properties of neutron stars~\cite{H2020}. 

The German Research Foundation's Collaborative Research Center "SFB 1225 ISOQUANT" and the Transregional Collaborative Research Center "CRC-TR 211" represent examples of successful funding that integrates theoretical approaches with experimental validation to study intensely interacting matter under extreme conditions~\cite{ISOQUANT:2019,CRC211:2020}. Notably, CRC-TR 211, with its transregional character and strong focus on heavy-ion physics, has had a particularly broad influence on shaping the field in Germany. These funding successes reflect a broader recognition that the most impactful advances in heavy-ion physics emerge from the interplay between theoretical development and experimental validation.

\subsection{Geographical Distribution of Contributions}

\begin{figure}[H]
\centering
\includegraphics[width=\textwidth]{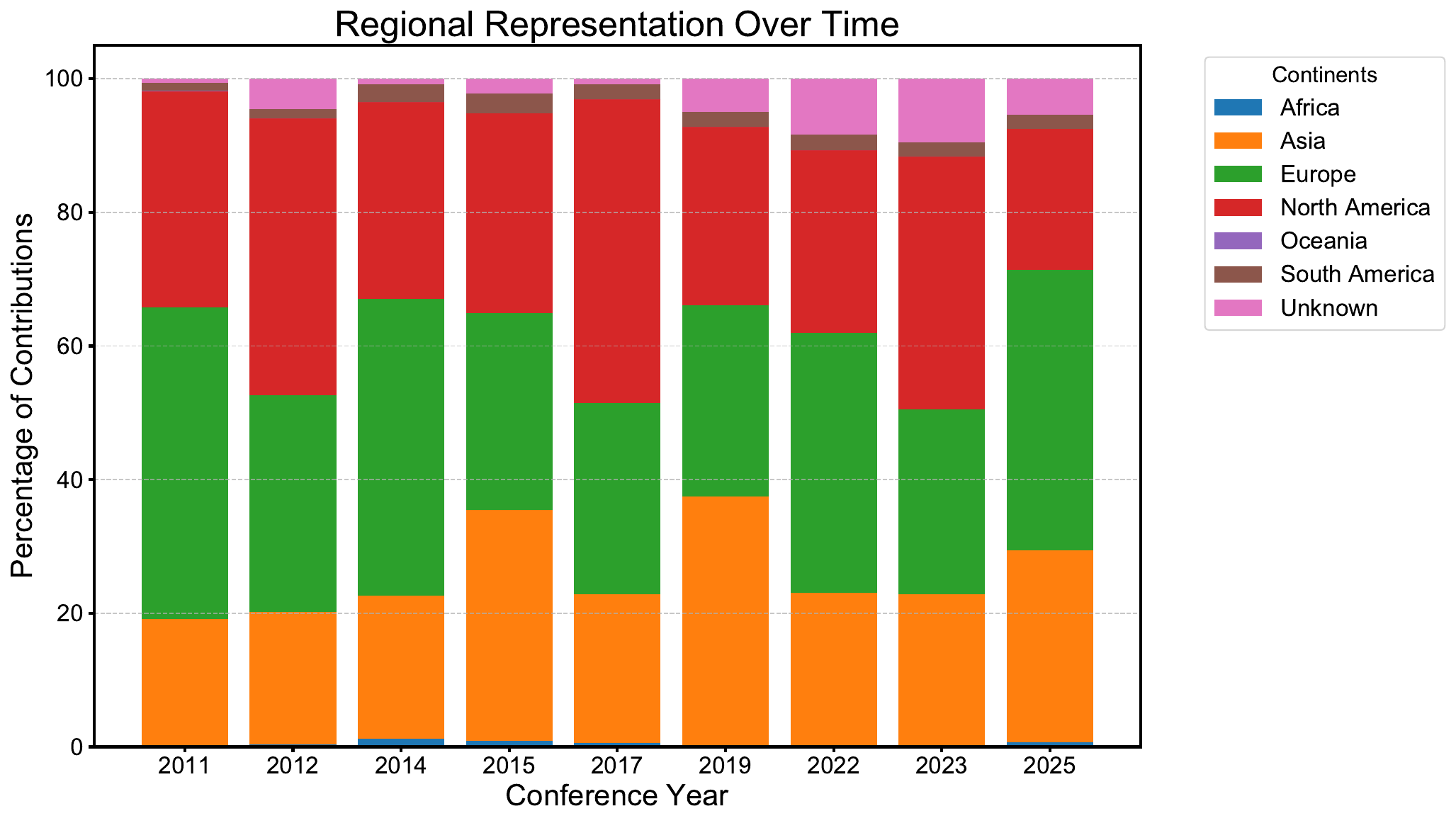}
\caption{Detailed view of regional representation at Quark Matter conferences from 2011 to 2025. This stacked area chart shows the percentage of total presentations from each geographical region for each conference year. This shows both persistent patterns and year-to-year fluctuations in international participation. North America and Europe consistently contribute the largest proportion of presentations, while Asia shows gradual growth in representation. Annual variations correlate with conference host locations, with regions typically showing elevated participation when hosting.}
\label{fig:regional_diversity_by_year}
\end{figure}

Figure~\ref{fig:regional_diversity_by_year} expands upon the regional analysis with a more detailed year-by-year breakdown.
The temporal trends show notable fluctuations that often correlate with conference locations. Conferences held in Europe typically show elevated European participation, and similar regional effects are observable for North American and Asian venues. This "host region effect" appears to be a consistent factor in participation patterns.

Asian participation shows a gradual upward trend across the analyzed period, reflecting the growing strength of heavy-ion physics research in countries like China, Japan, India, and South Korea. This trend suggests a slow but steady diversification of the field's geographical base, though predominance from research centers associated with RHIC (USA) and LHC (Europe) remains strong. The increasing Asian representation indicates both expanding experimental capabilities and theoretical expertise in the region, contributing to a more globally distributed research community.

This clearly identifies years with anomalous participation patterns, such as during the COVID-19 pandemic, when travel restrictions significantly impacted international representation at conferences. These disruptions highlight the sensitivity of international scientific participation to global events.

The minimal representation from South America, Africa, and Oceania persists throughout the period, with only marginal increases in certain years. This pattern highlights regions where focused efforts might be needed to increase participation and representation in the field.

\begin{figure}[H]
\centering
\includegraphics[width=\textwidth]{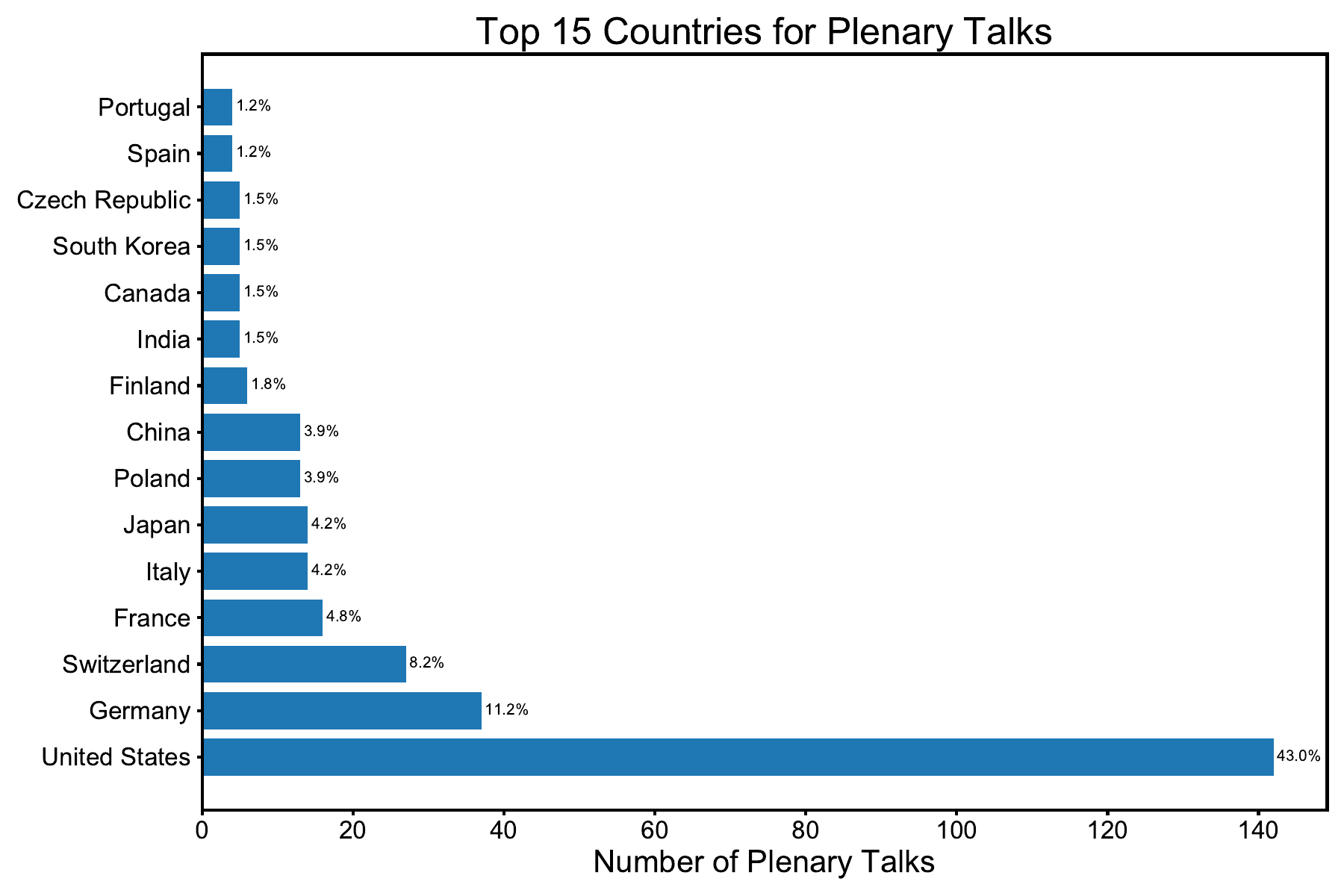}
\caption{Distribution of plenary talks by country across Quark Matter conferences from 2011 to 2025. The horizontal bars show the number of plenary presentations given by speakers from each of the top 15 countries, with percentage labels indicating relative contribution.}
\label{fig:country_plenary}
\end{figure}

Figure~\ref{fig:country_plenary} displays the distribution of plenary talks by country across Quark Matter conferences. Plenary talks represent the highest visibility presentations at these conferences and are typically allocated to highlight significant advances in the field.
There is a consistent dominance by a small number of countries, particularly the United States, which maintains a substantial share of plenary talks across all conference years. This reflects the significant investment in heavy-ion research infrastructure and personnel in the US, home to the RHIC facility and major ALICE, CMS, and ATLAS heavy-ion programs.
European countries collectively represent another major block, with Germany, France, and the UK consistently present. The distribution among European countries shows some variability between conferences, potentially reflecting both the location of the conference (European conferences tend to have more European speakers) and shifts in research output.
Asian representation, particularly from China, Japan, India, and South Korea, shows interesting dynamics over time. We observe a general trend of increasing representation from these countries, especially in more recent conferences, reflecting growing investment in the field in these regions.
The data reveal potential geographical imbalances in high-visibility speaking opportunities while some variation is expected due to differences in community size and research output.

\subsection{Parallel Talk Distribution}

\begin{figure}[H]
\centering
\includegraphics[width=\textwidth]{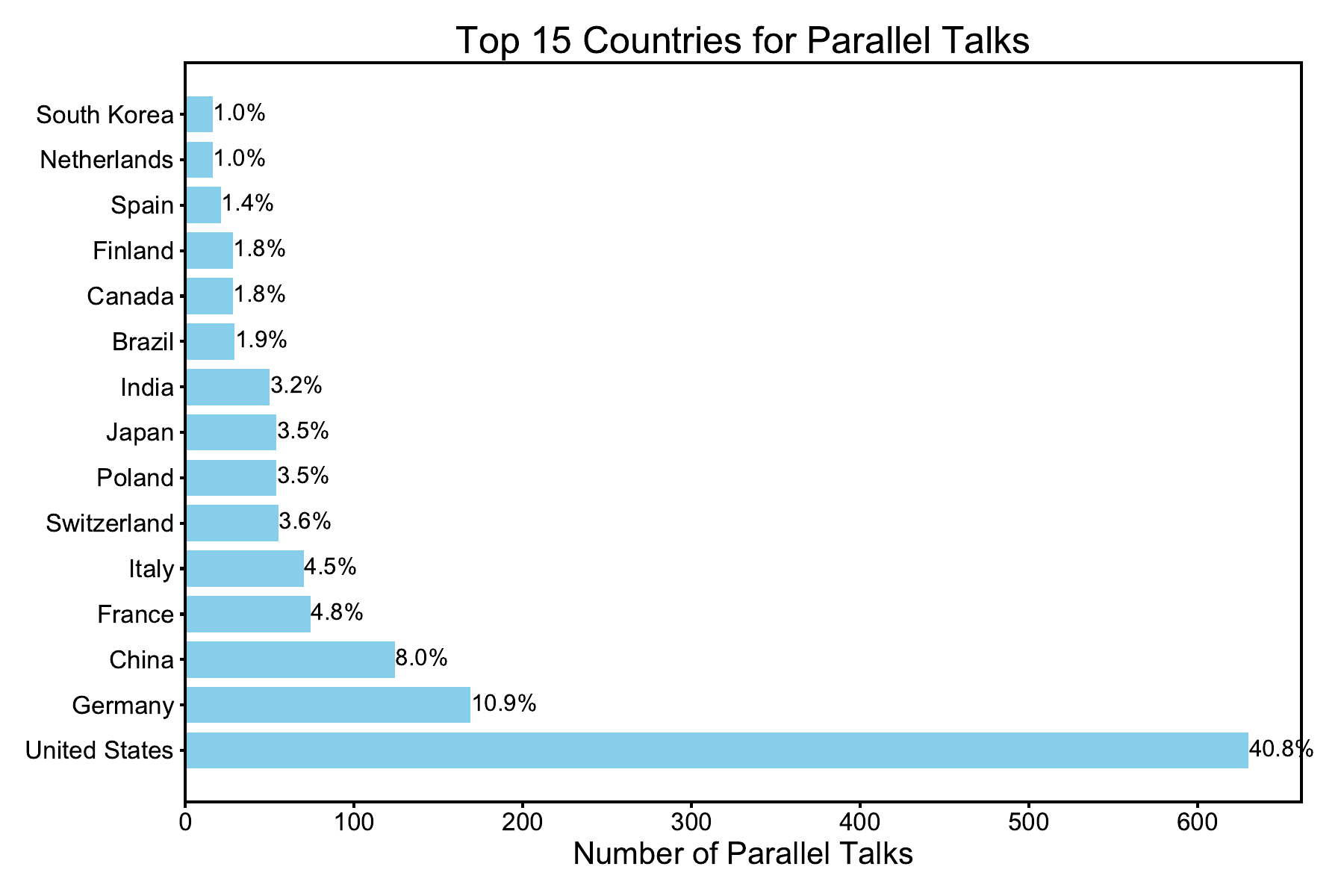}
\caption{Distribution of parallel talks by country across Quark Matter conferences from 2011 to 2025. The horizontal bars show the total number of parallel presentations from each country, with percentage labels indicating relative contribution.}
\label{fig:parallel_talks}
\end{figure}

Figure~\ref{fig:parallel_talks} presents the distribution of parallel talks across countries, providing insight into the broader participation patterns beyond plenary presentations. The distribution shows a clear hierarchy in participation levels, with a few countries contributing a large proportion of parallel talks. The USA shows the highest participation rate, followed by other major research centers. There is a long tail of countries with smaller but significant contributions, indicating broad international participation. The distribution pattern differs somewhat from that seen in plenary talks, suggesting different selection dynamics between presentation types.

\begin{figure}[H]
\centering
\includegraphics[width=\textwidth]{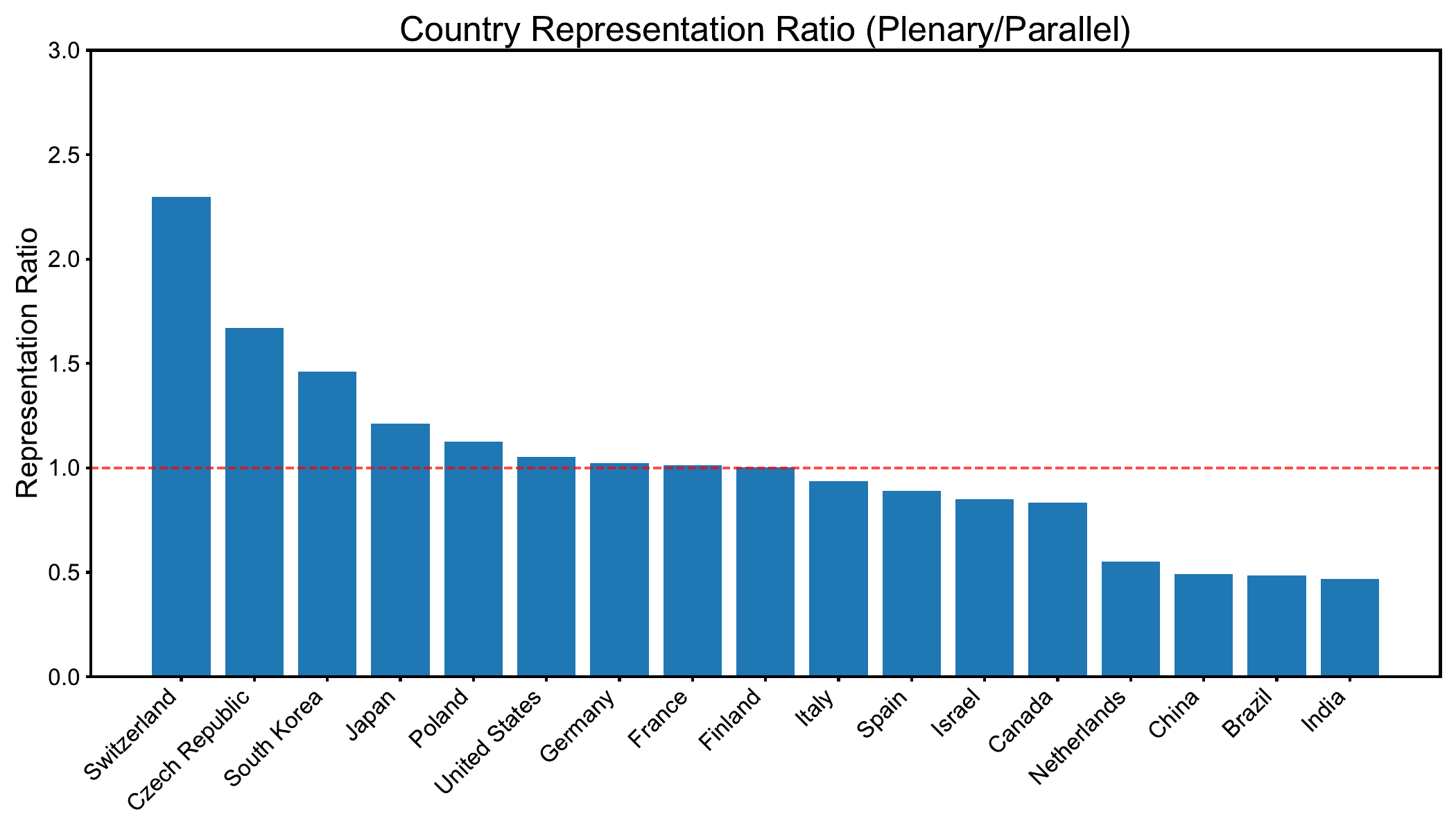}
\caption{Representation ratio between plenary and parallel talks by country. This chart compares each country's share of high-visibility plenary talks to their share of parallel sessions, with values above 1.0 indicating overrepresentation and values below 1.0 indicating underrepresentation. Only countries with at least 5 parallel talks are included to ensure statistical significance, and a reference line at ratio=1.0 marks perfectly proportional representation.}
\label{fig:representation_ratio}
\end{figure}

Figure~\ref{fig:representation_ratio} presents an measure of representational equity in Quark Matter conferences. The representation ratio is calculated as: Ratio = (Country's percentage of plenary talks) / (Country's percentage of parallel talks). Values above 1.0 indicate a country receives more plenary visibility than their parallel talk participation would suggest. Values below 1.0 indicate less plenary visibility than expected based on parallel talk participation. A ratio of exactly 1.0 represents perfectly proportional representation.
Here, only countries with at least 5 parallel talks are included to avoid statistical anomalies from small sample sizes. The implementation includes calculation of parallel and plenary percentages for each country, filtering to include only statistically significant countries, sorting countries by representation ratio, a reference line at 1.0 to highlight the threshold between over and underrepresentation, and a cap on the y-axis scale to prevent extreme outliers from distorting the visibility.

The results show substantial variation in representation ratios across countries. A small number of countries have notably high representation ratios ($>$1.5). A larger group of countries cluster around the proportional representation line (0.8-1.2). Many countries fall below the proportional representation threshold. The pattern suggests structural factors may influence plenary selection beyond research volume.

\begin{figure}[H]
\centering
\includegraphics[width=\textwidth]{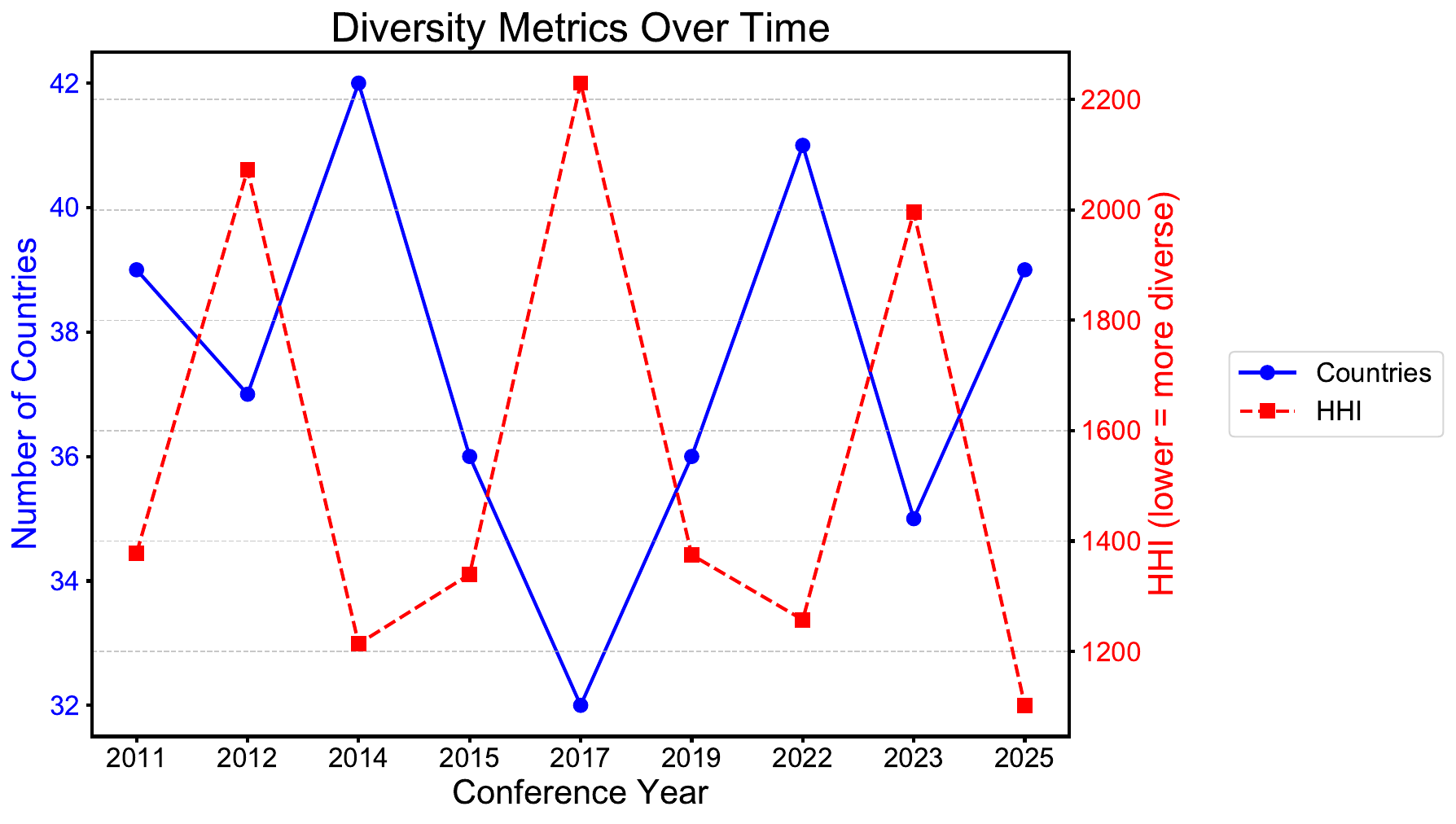}
\caption{Evolution of diversity metrics over time. The blue line shows the number of countries represented at each conference, providing a direct measure of geographical diversity. The red line displays the Herfindahl-Hirschman Index (HHI), a measure of concentration where lower values indicate more diverse representation. The complementary metrics together provide a picture of how the international composition of the conference has evolved.}
\label{fig:diversity_metrics}
\end{figure}

Figure~\ref{fig:diversity_metrics} tracks the evolution of diversity in Quark Matter conferences using two complementary metrics. The primary y-axis (blue) tracks the number of countries represented at each conference. The secondary y-axis (red) displays the Herfindahl-Hirschman Index (HHI), which measures concentration/diversity. The HHI is calculated as the sum of squared market shares: $\text{HHI} = \sum_{i=1}^{N} s_i^2$, where $s_i$ is the percentage of talks from country $i$, and $N$ is the total number of countries \cite{Rhoades1993}. Lower HHI values indicate more equitable distribution of talks across countries. Different marker styles distinguish between the two metrics for clarity.

The analysis reveals several key insights about the evolution of international diversity. HHI values show a declining trend in most periods, indicating greater equity in distribution of talks. Certain years show countertrends where diversity decreased despite conference growth. Conference location appears to influence diversity, with some correlation between host region and diversity metrics. 
While the country count measures breadth of participation, the HHI captures equitability of distribution. Together, they offer a better view than either metric alone could provide. This approach allows distinction between scenarios where many countries participate but with highly uneven distribution (smaller HHI values) versus scenarios with fewer countries but more balanced representation (closer to 1.0 HHI).

\subsection{Institutional Representation}

\begin{figure}[H]
\centering
\includegraphics[width=\textwidth]{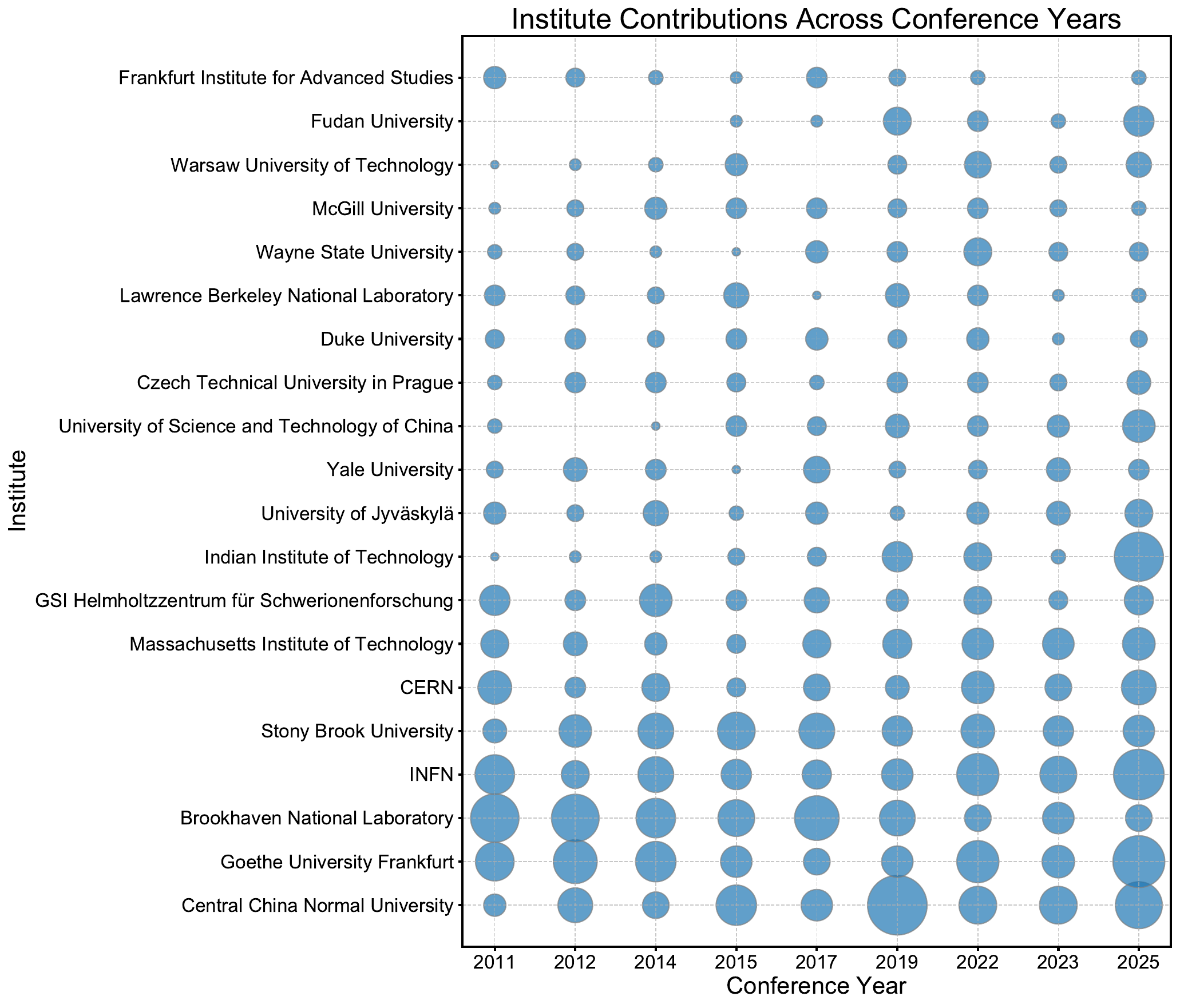}
\caption{Institute contributions across Quark Matter conferences from 2011 to 2025. This bubble chart displays the presence and contribution volume of the top 30 institutes across conference years, with bubble size proportional to the number of presentations.}
\label{fig:institute_bubble}
\end{figure}

Figure~\ref{fig:institute_bubble} provides a perspective on institutional participation patterns through a time-series bubble chart. In this chart, the X-axis represents conference years, the Y-axis represents institutes (sorted by total contribution), and bubble size represents the number of presentations in that year, including plenary, parallel, and poster contributions.

The chart reveals distinct participation patterns among institutes. Some institutes maintain a consistent presence with relatively stable contribution volumes across all conferences. Others show more intermittent participation, with strong showings in specific years followed by reduced presence. Major national laboratories show some of the most consistent participation patterns, while university participation appears more variable, potentially reflecting shifting research priorities or funding cycles. Some institutes show coordinated patterns, with increased participation in the same conference years.

\begin{figure}[H]
    \centering
    \includegraphics[width=\textwidth]{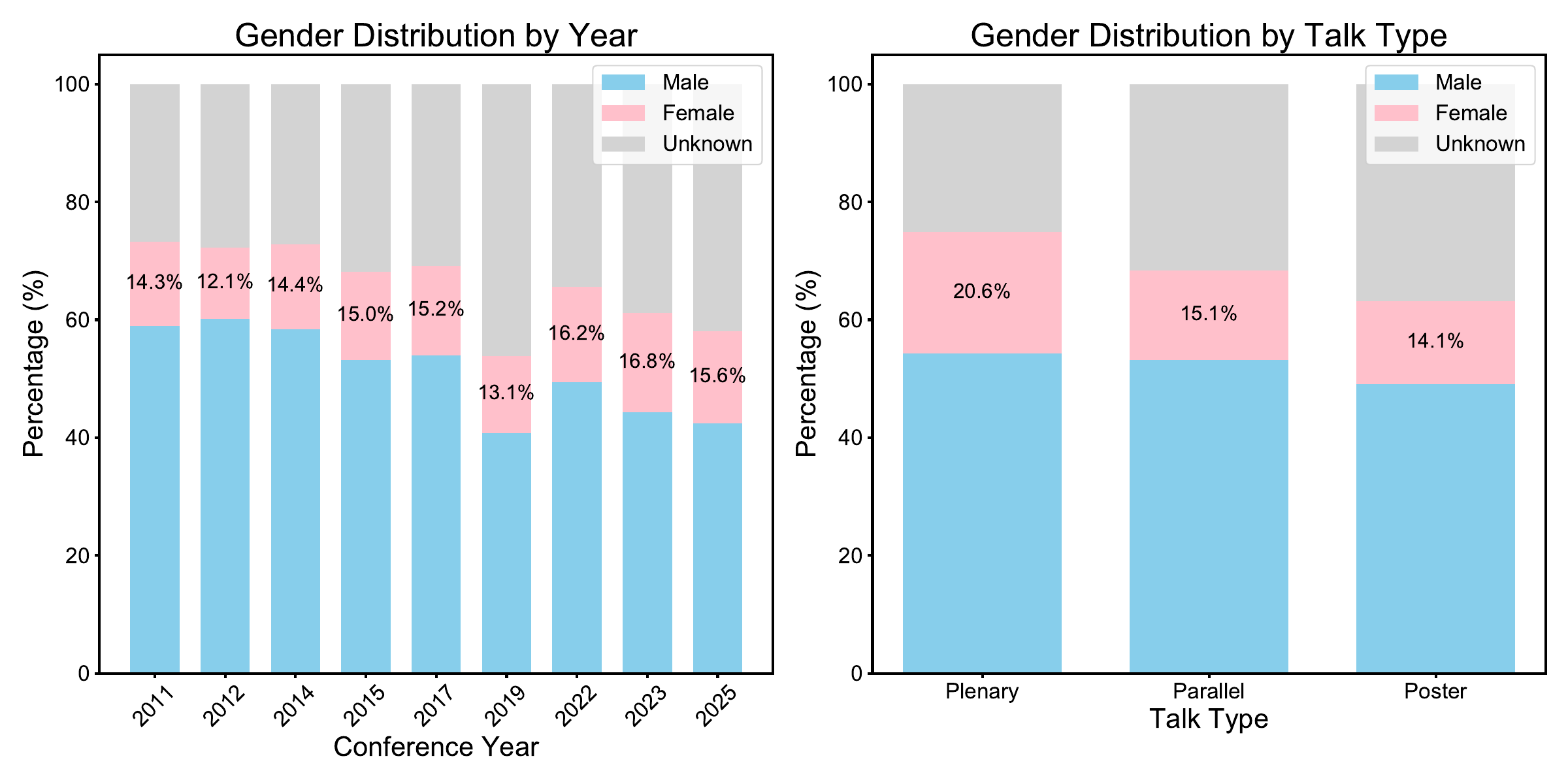}
    \caption{Gender diversity in Quark Matter conferences. The left panel displays the gender distribution by year, showing the percentage of male, female, and unknown speakers. The right panel presents gender distribution by talk type (plenary, parallel, and poster).}
    \label{fig:gender_diversity}
\end{figure}
    
Figure~\ref{fig:gender_diversity} examines gender diversity across Quark Matter conferences. The left panel shows gender distribution percentages over time (2011-2023). The right panel displayes gender distribution across different talk types. Stacked bar charts show relative proportions of male, female, and unknown presenters. Percentage labels highlight female representation directly on the figure.
With year-to-year fluctuations, female representation has shown a gradual increase over the analyzed period. The right panel shows interesting variations across presentation types, with plenary talks showing higher female representation than parallel and poster presentations. The "unknown" category accounts for a non-negligible percentage, representing cases where gender could not be identified.
    
Our gender analysis employs the gender-guesser library~\cite{gender-guesser}, which uses a database of names across different cultures to estimate gender from first names. Names classified as "female" or "mostly female" were categorized as female, while "male" or "mostly male" were categorized as male. Ambiguous or unrecognized names were labeled as "unknown." This approach has inherent limitations, as it cannot account for all cultural variations in naming or non-binary gender identities. The results should be interpreted as approximate patterns rather than definitive statistics. Despite these limitations, the observed patterns for the different talk types are consistent enough because of the randomness across the talk types. The "unknown" gender classification in our analysis is strongly influenced by the fact that conference registrants are not required to disclose their gender information. Additionally, while our analysis includes a category for non-binary gender identities, the actual representation of non-binary individuals in the dataset is very small, likely due to both the limitations of our name-based classification method and the voluntary nature of gender disclosure in conference registration.

\section{Discussion}
\label{sec:discussion}

This historical analysis of Quark Matter conferences illuminates the evolution of heavy-ion physics through its leading conference series. Our quantitative analysis provides insights into how these conferences have served and can better serve the scientific community.

The keyword analysis reveals how scientific focus has evolved over the past decade. We observe a transition from facility-focused research toward phenomenon-focused investigations, with increasing specialization and technical sophistication in later conferences. The emergence of new research directions—particularly investigations of collectivity in small systems and machine learning applications—demonstrates the field's evolution from establishing basic QGP properties toward detailed investigations of specific phenomena. These keyword dynamics highlight the field's responsiveness to experimental results, such as collective behavior in small collision systems.

Our geographical and institutional representation analysis reveals patterns reflecting structural factors and organizational efforts. While plenary presentations show some concentration among certain countries and institutions, this reflects differences in community size, research infrastructure, and historical development across regions. Quark Matter organizers implement layered selection processes, and voting reduces potential biases, though quantifying their effectiveness remains challenging due to variables like group sizes and funding levels.

The data show several practical approaches already implemented to enhance diversity: transparent selection criteria balancing geographical diversity with scientific excellence, flash talk programs for researchers from underrepresented institutions, systematic rotation of conference locations, and demographic data monitoring. While maintaining scientific excellence as the primary criterion, attention to regional representation has ensured broader participation.

The findings indicate a vibrant and evolving research community addressing increasingly complex questions. The shift from focusing on facilities to concentrating on phenomena highlights the field's maturity, along with the growing interactions between theoretical and experimental groups. While Europe and North America continue to have a strong presence due to established infrastructure, the data show gradual diversification, especially from Asian countries.

Conference organizers have addressed representation challenges through balanced selection criteria, flash talk programs, location rotation, and data tracking—demonstrating a commitment to inclusivity while maintaining scientific standards. In line with this commitment, Quark Matter conferences adhere to the International Union of Pure and Applied Physics (IUPAP) policies on the free circulation of scientists and harassment prevention. As stated on the official QM2025 website, IUPAP "actively upholds" the principle that scientific progress requires "freedom of movement, association, expression and communication for scientists" and "opposes any discrimination on the basis of such factors as ethnic origin, religion, citizenship, language, political stance, gender, or age" \cite{IUPAP2025}. Additionally, conference organizers appoint advisors to ensure all participants "enjoy a comfortable experience" and to address any harassment issues that might arise. These policies reflect the community's recognition that scientific excellence flourishes in environments that welcome diverse perspectives and ensure equitable participation.

This study demonstrates how data-driven analysis can provide valuable insights for scientific communities. Making participation patterns publicly available will help foster discussions about the future development of heavy-ion physics conferences.

\section*{Acknowledgments}
We thank the organizers of Quark Matter conferences for making the presentation data publicly available through Indico, enabling this analysis. We are grateful to colleagues who provided valuable feedback on our analysis methodologies and visualizations during the development of this work, especially to the students of Ultra-relativistic Heavy Ion Physics Lecture course (FYSS4551) in 2025 at the University of Jyväskylä, C. Sporleder, and P. Runko for their help with data refinement, result discussions, and manuscript feedback. We thank Kari Eskola for his valuable comments and insights. We are particularly grateful to the organizer of Quark Matter 2025, Dirk Rischke, and the organizers of the upcoming conference, In-Kwon Yoo and Sang-Yong Geon, for their constructive feedback and suggestions. Both authors acknowledge support from the Research Council of Finland through the Center of Excellence in Quark Matter (Project Nos. 346328 and 364193). We also acknowledge the use of Claude 3.5 Sonnet algorithm for assistance in developing ideas, refining the analysis, and debugging the code.

\bibliography{ref}

\appendix
\section{Quark Matter Conference Proceedings}
\label{app:proceedings}
This appendix provides a list of all Quark Matter conference proceedings from 1980 to 2025, serving as a reference resource for researchers seeking historical publications from the conference series.

\begin{table}[H]
\centering
\caption{Complete list of Quark Matter conference proceedings (1980-2025)}
\label{tab:proceedings}
\small 
\begin{adjustbox}{width=\textwidth} 
\begin{tabular}{clll}
\toprule
\textbf{No.} & \textbf{Year} & \textbf{Location} & \textbf{Proceedings} \\
\midrule
1 & 1980 & Berkeley, USA & LBL-Report 11123 \\ 
2 & 1982 & Bielefeld, Germany & World Scientific, 1982 \\ 
3 & 1983 & Brookhaven, USA & Nuclear Physics A418, 1984 \\ 
4 & 1984 & Helsinki, Finland & Berlin : Springer, 1985 - 305 p., https://cds.cern.ch/record/105394\\
5 & 1986 & Pacific Grove, USA & Nuclear Physics A461, 1987 \\ 
6 & 1987 & Nordkirchen, Germany & Zeitschrift für Physik C - Particles and Fields, Volume 38, 1988 \\ 
7 & 1988 & Lenox, USA & Nuclear Physics A498, 1989 \\ 
8 & 1990 & Menton, France & Nuclear Physics A525, 1991 \\ 
9 & 1991 & Gatlinburg, USA & Nuclear Physics A544, 1992 \\ 
10 & 1993 & Borlänge, Sweden & Nuclear Physics A566, 1994 \\ 
11 & 1995 & Monterey, USA & Nuclear Physics A590, 1995 \\ 
12 & 1996 & Heidelberg, Germany & Nuclear Physics A610, 1996 \\ 
13 & 1997 & Tsukuba, Japan & Nuclear Physics A638, 1998 \\ 
14 & 1999 & Torino, Italy & Nuclear Physics A661, 1999 \\ 
15 & 2001 & Stony Brook, USA & Nuclear Physics A698, 2002 \\ 
16 & 2002 & Nantes, France & Nuclear Physics A715, 2003 \\ 
17 & 2004 & Oakland, USA & J.Phys.G 30, 2004 \\ 
18 & 2005 & Budapest, Hungary & Nuclear Physics A774, 2006 \\ 
19 & 2006 & Shanghai, China &  Int.J.Mod.Phys.E 16, 2007 \\ 
20 & 2008 & Jaipur, India &  Indian J.Phys. 85, 2011 \\ 
21 & 2009 & Knoxville, USA & Nuclear Physics A830, 2009 \\ 
22 & 2011 & Annecy, France & J.Phys.G 38, 2011 \\ 
23 & 2012 & Washington DC, USA & Nuclear Physics A904-905, 2013 \\ 
24 & 2014 & Darmstadt, Germany & Nuclear Physics A931, 2014 \\ 
25 & 2015 & Kobe, Japan & Nuclear Physics A956, 2016 \\ 
26 & 2017 & Chicago, USA & Nuclear Physics A967, 2017 \\ 
27 & 2018 & Venice, Italy & Nuclear Physics A982, 2019 \\ 
28 & 2019 & Wuhan, China & Nuclear Physics A1005, 2021 \\ 
29 & 2022 & Krakow, Poland & Acta Phys.Polon.Supp. 16 (2023) \\ 
30 & 2023 & Houston, USA & EPJ Web Conf. Volume 296, 2024 \\ 
31 & 2025 & Frankfurt, Germany & EPJ Web Conf. TBD \\
\bottomrule
\end{tabular}
\end{adjustbox}
\end{table}

The proceedings of Quark Matter conferences represent a valuable historical record of the field's evolution. As shown in the table above, the publication venue for these proceedings has changed over time, with Nuclear Physics A serving as the primary journal for most conferences, while the Journal of Physics G has also published several volumes. This collection of proceedings provides researchers with access to over four decades of developments in heavy-ion physics, from the field's early theoretical foundations to the latest experimental results from RHIC and the LHC.

These proceedings capture the state of knowledge at each stage of the field's development and document the community's response to major experimental milestones, such as the first operations of the AGS, SPS, RHIC, and LHC heavy-ion programs. They also reflect the changing focus of research questions, methodological approaches, and theoretical frameworks that have shaped our understanding of the quark-gluon plasma and strongly interacting matter under extreme conditions.

For researchers interested in the historical development of specific topics within heavy-ion physics, these proceedings provide a chronological record that complements the publication of individual research papers in peer-reviewed journals. They also offer insights into the evolution of the field's priorities, methodologies, and collaborative structures that may not be apparent from journal publications alone.

\section{Quark Matter Conference Indico Pages}
\label{app:indico}

This appendix provides the complete list of Indico pages for Quark Matter conferences from 2011 to 2025 that were used for data collection in this analysis. These online repositories contain the presentation materials, abstracts, and participant information that formed the basis of our dataset.

\begin{table}[H]
\centering
\caption{Quark Matter conference Indico pages (2011-2025)}
\label{tab:indico_pages}
\small
\begin{adjustbox}{width=\textwidth}
\begin{tabular}{cll}
\toprule
\textbf{Year} & \textbf{Location} & \textbf{Indico URL} \\
\midrule
2025 & Frankfurt, Germany & \url{https://indico.cern.ch/event/1334113/} \\
2023 & Houston, USA & \url{https://indico.cern.ch/event/1139644/} \\
2022 & Krakow, Poland & \url{https://indico.cern.ch/event/895086/} \\
2019 & Wuhan, China & \url{https://indico.cern.ch/event/792436/} \\
2018 & Venice, Italy & \url{https://indico.cern.ch/event/656452/} \\
2017 & Chicago, USA & \url{https://indico.cern.ch/event/433345/} \\
2015 & Kobe, Japan & \url{https://indico.cern.ch/event/355454/} \\
2014 & Darmstadt, Germany & \url{https://indico.cern.ch/event/219436/} \\
2012 & Washington DC, USA & \url{https://indico.cern.ch/event/181055/} \\
2011 & Annecy, France & \url{https://indico.cern.ch/event/30248/} \\
\bottomrule
\end{tabular}
\end{adjustbox}
\end{table}

It should be noted that the data structure and availability varied across these Indico pages. Some conferences (particularly 2011, 2012, 2015, and 2023) had inconsistent JSON formats or incomplete participant information. For certain conferences, additional information about International Advisory Committees was obtained from conference-specific websites that are no longer accessible. These variations in data structure necessitated the development of robust parsing and data cleaning methodologies as described in the Data and Methods section.
\end{document}